\documentclass[aps,prd,epsf]{revtex4}
\usepackage{epsfig}

\newcommand{\be}{\begin{equation}}
\newcommand{\ee}{\end{equation}}
\newcommand{\bea}{\begin{eqnarray}}
\newcommand{\eea}{\end{eqnarray}}

\newcommand{\nn}{\nonumber}

\begin{document}


\title{\bf Bethe-Salpeter study of radially excited vector quarkonia}

\author{V. \v{S}auli}

\email{sauli@ujf.cas.cz}
\affiliation{Department of Theoretical Physics, Institute of Nuclear Physics Rez near Prague, CAS, Czech Republic  }

\begin{abstract}
We solve the Bethe-Salpeter equation (BSE) for a system of a heavy quark-antiquark pair  
interacting with a Poincare invariant generalization of screened linear confining potential. 
In order to get reliable description the  Lorentz scalar confining interaction is complemented by the effective one gluon exchange. Within presented model we reasonably reproduce all  known radial excitations of the vector charmonia.
We have found that $J/\Psi$ is the only charmonium left bellow naive quark-antiquark threshold $2m_c$, while the all excited states are situated above this threshold. We develop  a method which is enable to provide solution of full four dimensional BSE for the all  excited states. We discuss the consequences of the  use of the free propagators for calculation of excited states above the threshold. The Bethe-Salpeter string breaking scale $\mu\simeq 350MeV$  appears to be relatively larger then the one defined in various potential models $\mu\simeq 150MeV$.  
\end{abstract}

\maketitle


%
%
%
%
\section{Introduction}

Excited meson spectroscopy is a keystone experimental output, which
is essential for understanding of quark-antiquark interaction. 
The dynamics of meson constituents is driven by solely known strongly interacting quantum field theory-
quantum chromodynamics.  The explanation of quarks and gluons  confinement
 is one of the great challenge of theory of strong interaction. The same confining forces are responsible for           
a large degeneracy which emerges from the spectra of the angularly and radially excited resonances.
As reported in \citep{Bugg:2004xu,Afonin:2006vi,Glozman:2007ek,fig1afon} such degeneracy is observed in $p \bar p$ annihilation by the Crystal Ball collaboration at LEAR 
in BERN \citep{Aker:1992ny}. Similar can be deduced  from heavy quarkonia production in $e^{+}e^{-}$ anihilations.
Recently, BABAR, Belle, BESS and LHCb  experiments continue collection of various meson experimental data. 

 Following ideas of Ref. \citep{Wilson:1974}, confinement in heavy flavor  sector has typically been associated with a linearly rising potential between constituents \citep{Eichten:1978}. The spin degeneracy observed in heavy quarkonia spectrum  tell us that 
the main part of confining interaction should be largely spin independent. A various lattice fits leaded to various predictions for the potential between quark-antiquark states. A well known is  Cornell parameterization of the static Wilson loop derived potential \citep{Bali2001,Greensite2003},
\be \label{pot}
V(r)=-\alpha/r+\sigma r
\ee
which appeared to be suited for description of the first few excited mesons.

Recently in Refs. \citep{Licha2009,Chao1,Vento2011} it has been found that the meson spectroscopy is better described  by "confining"  potential which is bounded from above. While in the absence of dynamical quarks the nonrelativistic static potential can eventually  show up a linear asymptotic,
it should be flattened in large distance due to the string breaking associated with light hadron productions. In this respect, the exponential potential can be regarded as a screened version of the linear potential. The screening effect should be universal, e.g. scheme and gauge independent property of low energy QCD as  the creation of the light quark-antiquark pairs is energetically favorable and pions and other light mesons is observable fact. The screening could be included in order to explain observed hierarchy of  radially excited heavy mesons. Obviously, the spectrum deviates from the linear Regge trajectory, for a proposal of 5s and 6s charmonium candidate see \citep{RUPP}, however recall, the deviation from the linear Regge trajectory is expected in the light meson sector as well \citep{PETR}.    

The Coulombic part of the potential is naturally expected in QCD, however as shown recently, it can be suppressed in charmed meson sector by not yet understood mechanism  \citep{najit}. Furthermore, it is related with gluon exchange  then screening of Coulomb potential is expected as well.
This a bit more subtle matter could be related with soft gluon mass generation \citep{PBC1,PBC2} through the Yang-Mills Schwinger mechanism. Actually the dynamical gluon mass  generation is  suggested by finiteness of the lattice  gluon propagator in the deep infrared, for the recent lattice data on gluon propagator in Landau gauge see \citep{Mendes}. We expect, the screening mass characterizing string breaking and the soft gluon mass have similar size $\simeq \Lambda_{QCD}$.

For bound states which lie above the naive quark-antiquark threshold, the BSE kernel becomes singular, which causes that  many usual numerical treatments fails and/or become impossible in practice.
To authors knowledge,  plethora  of solutions have been obtained  in relativistic quantum mechanic \citep{RQM} or by solving  various phenomenological 3D reduction of BSE \citep{TJON1993,INSTANT}, however comparison with the original BSE is completely missing. 
The way of  three-dimensional reduction of the BSE is rather arbitrary and difficult to control without keeping the solution of the original BSE. Actually, choice of 3d equation has a large effect already  for the ground states.  In the study \citep{TJON1993} $200 MeV$ difference between Equal Time (ET) and certain  Quasipotential Equation has been found for the charmonium ground state $l=0$. This increases slightly for higher $l$ as well. The differences between  Schroedinger  and 3d approximated  BSE spectra  appears to be more significant for higher excited states, where following ref. \citep{TJON1993} the difference has been estimated $M_{QM}-M_{ET}\simeq 300MeV$ for $l=4,5$ charmonia.  The comparison between  various relativistic approximations has not been studied for radially excited states, however we expect very similar effect there.  From all this is apparent that relativistic covariance is important for charmonium and it will be useful to have the solution without any 3d approximation.  Due to this reasoning, using  single component approximation, but keeping the full four dimensionality of   BSE we study the effect of retardation in the slopes of radially
excited vector mesons. The experimental knowledge of heavy vector quarkonia is the main reason
why we concern spin 1 mesons as a first (as time is being the  pseudoscalars have been computed already in \citep{SAULI2}).

\section{Kernel for Bethe-Salpeter equation}

 In the Quantum Field Theory the two body bound state is described by 
the three-point bound state vertex function, or equivalently by 
the BS amplitude. Both of  them are solutions of the 
corresponding  covariant four-dimensional 
BSE  \citep{BETHE}. 
In principle, common framework of Schwinger-Dyson and BSEs  offers  unique  Poincare invariant  generalization of quantum mechanical picture sketched above, however a practical solutions are always incomplete due to the truncation of the equations system. 
Due to this fact we rather phenomenologically estimate what should be the form of the  BSE kernels here. 
In this paper we  use  "confining" interaction kernel of the form
\be  \label{kernel}
V_s(q)=\frac{C}{(q^2-\mu^2)^2} \, ;
\ee
which is certain screened form of   $\simeq 1/q^4$ scalar interaction.
The remaining considered part of the interaction kernel has the Dirac decomposition
identical with the one gluon exchange $\simeq \gamma^{\mu}_{\alpha\beta}V_v\gamma^{\nu}_{\alpha^{,}\beta^{,}}$.  

For clarity we write down the BSE completely here

\bea \label{BSEcelkove}
S^{-1}(q+P/2)\chi(p,q)S^{-1}(q+P/2)&=&-i\int \frac{d^4k}{(2\pi)^4}\gamma_{\mu}\chi(k,P)\gamma_{\nu}G^{\mu\nu}(k-q)
\nn \\
&-&i\int \frac{d^4k}{(2\pi)^4}\chi(k,P)V_s(k-q) \, ;
\nn \\
G^{\mu\nu}(k-q)&=&g^{\mu\nu}V_v=\frac{g^2g^{\mu\nu}}{(k-q)^2-\mu_g^2}  \, ; 
\eea

Two scalar functions $V_s$ and $V_v$ in (\ref{BSEcelkove}) completes our simple Poincare invariant generalization of quantum mechanical potentials.
Here, clearly $G^{\mu\nu}$ represents effective gluon propagator in Feynman like gauge, where the effective soft gluon mass $\mu_g$ has been introduced.  
The double pole scalar interaction $V_s$ leads to regular exponential potential in the position space.  
Actually,  in heavy quark limit one can consider three dimensional potential

\bea
V_s^{QM}(\vec{k}) 
&=& \int_0^\infty  dr  \, { 4 \pi r \sin( k r ) \over k }  \, V( r)  
\nonumber \\
&=&  \sigma { - 8 \pi \over ( {\bf \vec{k}^2} + \mu^2)^2   } \ , 
\eea
where the potential in position space reads 
\be
V(r)=- \sigma {e^{- \mu \, r} \over \mu} \, .
\label{mechanic}
\ee

Thus in certain sense the BSE model considered in this paper  represents  relativistic generalization of  the models considered in  
\citep{Chao1,Licha2009}.
 $S$ in Eq. (\ref{BSEcelkove}) stands for charm quark propagator, which in our  simplest approximation is taken as
\be
S^{-1}(l)=\not l-m_c\, , 
\ee  
and $\chi_V$ represents  the Bethe-Salpeter wave function which  has the general form:
\bea
\chi_V(q,P)=\not\epsilon \chi_{V0}+ \not P \epsilon.q\chi_{V1}+ \not q \epsilon.q\chi_{V2}+\epsilon.q\chi_{V3}
\nn \\
+ [\not\epsilon,\not P]\chi_{V4}+[\not\epsilon,\not q]\chi_{V5}+[\not q,\not P]\chi_{V6}+i\gamma_5\not t\chi_{V7} \, \, ,
\eea
with $t_{\mu}=\epsilon_{\mu\nu\alpha\beta}q^{\nu}P^{\alpha}\epsilon^{\beta}$,
$\epsilon^2=-1$, $\epsilon.P=0$.

Munczek and Jain \citep{JAMU} have shown that $V0$ component is   dominant for the all ground state  mesons, which  dominance is particular for mesons made form  heavy flavor (anti)quarks. The same is valid for the case of all  pseudoscalar radial excitations  \citep{sauli},  which strongly argue  for that if there  is a  dominant component  it stay to be  dominant one for higher excited states independently on the meson spin.
Therefore we assume the same apply for excited vectors here,  and we neglect all other components in  presented study.

Within the approximation the Bethe-Salpeter equation in the rest frame reads
\be \label{BSE1}
\chi_{V0}(p_E,P)=\frac{-p^2_E-m^2-P^2/4}{(-p^2_E-m^2+M^2/4)^2+q^2_4M^2} I_0 \,\, ,
\ee
where
\bea
I_0&=&-\int\frac{d^4k_E}{(2\pi)^4}\left[-2V_v+V_s\right] \chi_{V0}(k_E,P)  \, \, ;
\nn \\
V_v&=&\frac{g^2}{q^2_E+\mu^2} \, ;
\nn \\
V_s&=&\frac{C}{(q^2_E+\mu^2)^2} \, ,
\nn \\
\eea
where we have performed Wick rotation into the Euclidean space.  For Euclidean momenta we use the convention $k_E=(k_4,\vec{k})$, $k_E^2=k_4^2+{\bf k}^2$, while the total momentum is kept timelike $P^2=-P^2_E=M^2$ as required for the bound states.
An extra  problem arises for bound states which are heavier then sum of constituents quark masses. 
As we are using propagators with single real poles, the BS wave function becomes singular since it is  proportional to  the product of the quark propagators.
For the vertex function, the  threshold like singularity should  appear for the solution with $P^2>4m_q^2$. Due to this fact the numerical integration requires special numerical care and we found very advantageous to define the following auxiliary function
\bea \label{regul}
A(p_E,P)&=&\chi_{V0}(p_E,P) G_A(k,P) \, ,
\nn \\
G_A(k,P)&=&\left[\frac{p^2_E+m^2+P^2/4}{(-p^2_E-m^2+M^2/4)^2+p^2_4M^2+\epsilon^4}\right]^{-1}
\eea
where $\epsilon $ is a small regulator mass satisfying $\epsilon^4<<p^2_4M^2$. Within the numerics the regulator can be  limited to zero. In limiting case  we should add the residuum contribution stemming  from the propagator pole. In the presented study we keep the regulator always finite and we neglect the residuum as well. It avoids complicated analytical continuation to the Minkowski space in this case.  

Finally we integrate over the angles of 3d momentum  subspace.
Explicitly written the BSE (\ref{BSE1}) reads 
\be
A(p_E,P)=\int_{-\infty}^{+\infty}\frac{dk_4}{(2\pi)} \int_{0}^{\infty} d{\bf k} {\bf k}^2 A(k_E,P) G_A(k,P)\int \frac {d\Omega_{3d}}{(2\pi)^3}\left[-2V_v+V_s\right] 
\ee
which can be finally rewritten as
\bea \label{BSEEuclid}
A(p,P)&=& \int_{-\infty}^{\infty}{d k_4} \int_{0}^{\infty}d {\bf k}\, 
A(k,P)G_A(k,P) \left[K_s+K_v\right] \, ;
 \\
K_s&=&\frac{C}{(2\pi)^3}\frac{\bf{k^2}}{\left[k^2_E+p^2_E-2k_4p_4+\mu^2\right]^2-4\bf{k^2}\bf{p^2}} \, ; 
\nn
\\
K_v&=&-\frac{g^2}{(2\pi)^3}\frac{\bf k}{\bf p}
\ln\left[\frac{k^2+p^2-2k_4p_4-2{\bf k}{\bf p}+\mu^2}{k^2+p^2-2k_4p_4+2{\bf k}{\bf p}+\mu^2}\right] 
\nn \, \, ,
\eea
where the functions $K_{v,s}$ stem from $\Omega$ integration:
\be
-2\int \frac{d\Omega_{3d}}{(2\pi)^4}V_v=K_v \, ,
\int \frac{d\Omega_{3d}}{(2\pi)^4}V_s=K_s \, .
\ee

\section{Numerical solution of the charmonium BSE}

It is well established that in  unconfining  theory a  bound states spectra obtained through the BSE and through the corresponding Schroedinger equation mutually agree  up to the small relativistic correction. 
In opposite,  the solution of BSE for excited states, which lie above constituents particle threshold represents rather difficult numerical problem and no comparison is known in the literature, at least to the author.

Due to the presence of the kernel singularity more or less standard matrix methods \citep{BLAKRA2010} fail  since the inversion of numerical matrices is not possible.  Also we do not explore more or less conventional  expansion into the orthogonal polynomials which loses its efficiency  when, as one expects, relatively large number of polynoms is necessary. Instead of, we rather solve the full two dimensional integral equation by the method of simple iterations.
For this purpose we discretize $P^2$ and step by step we are looking for the solution of the BSE with given $P_i^2$. 

The BSE for bound states is a homogeneous integral equation and it satisfies usual normalization condition. Instead of  using this, 
to achieve a good numerical stability of the iteration process we implement normalization condition through an auxiliary function $\lambda(P)$ and solve the following equation:  
\bea \label{BSEEnum}
A(p,P)&=&\lambda(P) \int{d k_4} \int d {\bf k} A(k,P) G_A(k,P) K_E \, .
\eea

The following  has been found as  a particular useful choice for the function  $\lambda(P)$ 
\be
\lambda^{-1}(P)=\int{d k_4} \int d {\bf k} A(k,P)^2 f(k,P) \, ,
\ee
where  arbitrary positive weight function $f$ was chosen to be Gaussian in ${\bf k}$ and $k_4$.
Implementation of such $\lambda(P)$  makes BSE nonlinear but mainly numerically stable.
 Clearly the BSE solution has been identified when $\lambda(P)=1$ and when the difference between  consecutive iterations  vanishes at the same time. We found that these two conditions happen simultaneously, while for other values of  parameters $P,\lambda(P)\neq 1$ the numerics do not provide vanishing difference between iterations.

\begin{figure}[t]
\begin{center}
\centerline{  \mbox{\psfig{figure=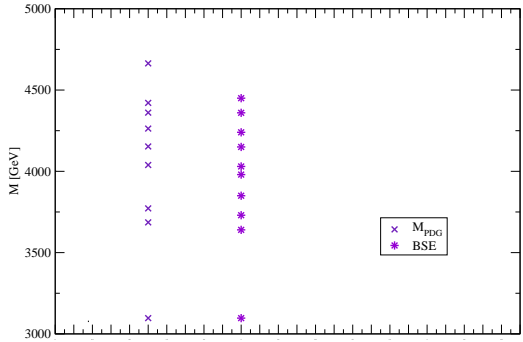,height=8.0truecm,angle=270}} }
\label{regge}
\caption{Comparison of BSE solution with PDG data for vector charmonium.}
\end{center}%
\end{figure}


   Numerical convergence of $\lambda\rightarrow 1$  depends on the density of the integration points in the important domain of momenta.    Recall here that even for 1-dim reduced BSE see \citep{ALKOFER} relatively large number of integration point was required in order to get precise solution. 
In presented study we are dealing with 2-dim integral equation with principal value integration, thus to achieve the same acuracy it necessarily   enlarge  number of integration points. Compromising between computational time and an estimated numerical error we are satisfied with the use of maximum $N_{k_4}*N_{\bf k}=184*96$ Gaussian points. In order to ensure the numerical stability we varied integration volume (in p-space) and the number of points as well. In order to get optimal density of integration points the upper boundary $\Lambda$ is introduced in momentum integrations.  More precisely, for  the variable ${\bf k}$ we map single integration interval $(0,1)$ into  $(0,\Lambda)$ by simple rescaling (and   $(-1,1)$ into  $(-\Lambda,\Lambda)$   for variable $k_4$).
To assure convergence $\lambda\rightarrow 1$ the density is varied such that for lower states smaller $\Lambda$ is used, while highly excited states are better obtained within higher $\Lambda$, with some cost of numerical precision in the later case.
The numerical  dependence on $N$ is shown in the Tab. 1. We also compare with the experimental data  in the Tab. 2 and Fig. 1. 
For purpose of brevity we  do not discussed all numerics here  leaving more detailed discussion 
elsewhere.  However, we can mention that as a numerical test we have checked  the  numerical method against the  scalar models \citep{SAULI}, where the resulting BSE spectra have been already obtained by different (and in fact  more accurate) method. Thus we solve 2dim BSE on a grid given by the component of relative momentum $k_4$ and ${\bf k}$ within the procedure described above, which  provides the spectra within $\simeq 1/N\%$ accuracy ($N$ is number of points  in the single, say $\bf k$- integration). As the regulator $\epsilon$ is implemented here, we expect the accuracy  of excited vectors spectrum studied here could be comparable with the one of scalar models solutions. In our case we estimate the 1 $\%$ numerical error in determination of bound state masses.
Such numerical error  is  considerably smaller then  the expected amount of energy shift due to the open threshold effects ( effect associated with $D$ meson production, not  be confused with unphysical quark-antiquark threshold. An open charm is impossible to incorporate in the formalism of homogeneous BSE and it is completely ignored here). 

For clarity  we restrict ourself to the dynamics of charmonia in this paper. The numerical values of the models are the following
\bea 
C&=&5.418 GeV^2\, \, ; \alpha_s=g^2/(4\pi)=0.2 \\ 
\nn m_c&=& 1.5615 GeV ; \mu=\mu_g=364.35 MeV  \, 
\eea
 and we use the experimental  $J/\Psi$ mass  to fine tune the correct scale at the end  (in units where $m_c=1.5 \, ;\,  C=5 $ we have $M(J/\Psi )=2975$).     
In order to improve the   numerical stability the infrared regulator in (\ref{regul}) was adjusted as $\epsilon=0.03 GeV$ (for $\Lambda=1000 GeV$, $\epsilon$  value is taken slightly larger
then the smallest interval between integration points of the variable $ k_4$).
\begin{table}[t]
\begin{center}
\small{
\begin{tabular}{||c|c|c||c|c|c||c|c|c||c|c|c||} \hline \hline $M(32) $ &
$\lambda(32)$  & $\sigma(32) $ & $M(58)$
& $\lambda(58) $ & $\sigma(58) $ & $M(72) $ &
$\lambda(72)$  & $\sigma(72) $ & $M(96)$
& $\lambda (96) $ & $\sigma (96)$ \\
\hline \hline
3.717 & 0.995 & 1.2E-06 & 3.642 & 1.398 & 0.006 & 3.639 & 1.444 & 0.008 & 3.636 & 1.458 & 0.008 \\
\hline
3.832 & 1.015 & 1.5E-05 & 3.746 & 1.174 & 0.001 & 3.745 & 1.200 & 0.002 & 3.733 & 1.223 & 0.002 \\
\hline
3.914  & 0.994 & 2.0E-06 & 3.865 & 0.998 & 2.5E-07 & 3.867 & 1.000 & 1.7E-09 & 3.855 &  1.017&  1.8E-05 \\
\hline
4.040  & 0.993&  2.5E-06 & 3.999 & 1.027 & 4.5E-05 & 3.996 & 0.979 & 2.6E-05 & 3.980 & 0.985 & 1.23E-05 \\
\hline
4.151 & 0.983 & 1.7E-05 & 4.061 & 1.020 & 2.5E-05 & 4.049 & 0.983 & 1.7E-05 & 4.033 & 1.01 & 2.24E-05 \\
\hline
4.289 & 1.012&  1.0E-05 & 4.177 & 1.024 & 3.7E-05 & 4.168 &  1.002 & 3.5E-07 & 4.149 &1.00 & 5.24E-06 \\
\hline
4.458 & 1.024 & 3.7E-05 & 4.277 & 1.030 & 5.7E-05 & 4.262 &  1.002 & 3.5E-07 & 4.243 & 1.01 & 2.03E-05 \\
\hline
4.604 & 1.025 & 4.0E-05 & 4.402 & 1.013 & 1.1E-05 & 4.340 &  0.977 & 3.1E-05 & 4.365 &1.02 &1.3E-05 \\
\hline
4.840 & 1.017 & 1.8E-05 & 4.549 & 0.992 & 3.5E-06 & 4.530 & 0.980 & 2.4E-05 & 4.503 & 1.02 & 1.4E-05 \\
\hline
5.008 & 1.021 & 1.8E-05  & 4.690 & 0.998 & 1.4E-07 & 4.671 & 1.007 & 3.98E-06 & 4650 & 1.005 & 5.1E-06 \\
\hline
5.318 & 0.989 & 7.00E-06 & 4.890 & 0.99  & 1.4e-07 & 4871  & 1.01  & 2.07E-05 & & & \\
\hline
5.518 & 1.009 &  5.5E-06 & 5.055 & 0.998 & 1.4E-07 & 5.030 & 1.009 & 5.09E-06 & & & \\
\hline
5.934 & 1.018 & 2.1E-05  & 5.337 & 1.013 & 1.0E-05 & 5.308 & 1.01  & 1.20E-05 & & & \\
\hline
6.171 & 1.015 & 1.5E-05  & 5.527 & 0.987 & 9.8E-06 & 5.496 & 1.02  & 2.60E-05 & & & \\
\hline
6.728 & 0.998 & 1.6E-07  & 5.909 & 0.97  & 3.5E-05 &       &       &          & & & \\
\hline
- & -&  -                & 6.140 & 1.00  & 2.5E-06 &       &       &          & & & \\
\hline \hline
\end{tabular}}
\caption[99]{Spectrum obtained for various number of mesh points $N=32,58,72,96$ used  at each single integral. 
The density of mesh point is regulated by $\Lambda=1000 GeV$ and infrared regulator$ \epsilon=0.03GeV$ is used, $\lambda$ and the iteration error $\sigma$ are displayed for completeness.
There is one more state observed at $3.880  \,(0.998 ,\,  2.5E-07)$ for N=58, which is skipped in the table  for better comparison. $M_{J\Psi}=3098$
is used as a fit.
\label{tab_bse1}}
\end{center}
\end{table}
\begin{table}[t]
\begin{center}
\small{
\begin{tabular}{|c|c|c|c|} \hline \hline $M_{96} $ &
$PDG$  & $n,l $   \\
\hline \hline
3097 & 3097 & 1s \\
\hline
3640   & 3686 & 2s \\
\hline
3730 & 3772 & 1d\\
\hline
3850 & -- &  \\
\hline
3980 & -- &  \\
\hline
4030 & 4039 & 3s \\
\hline 
4150 & 4153 & 2d\\
\hline 
4240 & 4263 & 4s\\
\hline
4360 & 4361 & 3d \\
\hline
4450 & 4421 & 5s \\
\hline \hline
\end{tabular}}
\caption[99]{Comparison with PDG data (second column) and calculated spectrum.
Quantum numbers correspond with  assumed quantum mechanical assignment \citep{Vento2011}.
\label{tab_bse2}}
\end{center}
\end{table}

As one see from the Table I, numerically $\lambda\neq 1$ for the first two states. In fact $\lambda$ do not cross  unit value, instead it posses positive local minimum which is stated in the Table. Further, when comparing with the experimental data, there are two more BSE solutions with masses in between $\psi^{''}$ and $\psi^{'''}$, while  the rest of calculated excited states quite nicely agree with the data. We have not found simple way (by varying parameters $C,\alpha,mu,m_c$) to exclude these two additional states from the solutions and we suggest the reason  why they are here in the Section bellow.

\section{Conclusion and discussion of the results}

We have formulated Lorentz covariant model for vector quarkonia, which is based on the BSE with phenomenological kernel.
With  aforementioned exception of two additional states, the resulting spectrum is comparable with the experiments whenever the experimental data are available.  The agreement between our results for higher states and the one measured in the experiments is impressive, remaining   difference between theory and experiments is due to the approximations, e.g.  due to the interplay of quark and $D,D*$ mesonic degrees of freedom, such  couple channel effects  are difficult to incorporate into the Bethe-Salpeter analysis presented here. 

The ground state is  situated  near bellow the  naive quark-antiquark threshold, while the all  excited states lie  above this  threshold.
We argue two more states which appear are  the artefact of inappropriate usage of free quark propagators. 
The question of confinement is beyond scope of the presented work, however we expect some  changes when confinement is correctly incorporated.
First of all, we expect the  quark propagators should not have a free particle pole and therefore the BSE could not posses ordinary threshold singularity in this case.  
In the paper we are dealing with BSE where the propagators  describe free -instead of confined- quarks. Therefore the threshold singularity unavoidably appears as an artefact here.
The lowest lying excited charmonia are the one  closest to the naive quark-antiquark threshold and we naturally must expect
some defect in the calculated spectrum.  Here, very likely it leads to mentioned appearance 
of two more excited states that we cannot find in the PDG, (for a recent attempt to find more definite answer see \citep{sauli}).

The model presented is very simple: The BSE kernel  consists  from vectorial effective one gluon exchange  and from the scalar infrared enhanced -double pole- interaction. A possible vector-scalar admixture of "confining" interaction has not been considered and we expect it must be small in order to suppress large hyperfine splitting. The string breaking mechanism is incorporated through the screening mass $\mu$ which is found to be  comparable to $\Lambda_{QCD}$ in our BSE study. Presence of Lorentz scalar interaction complemented by vector-vector interaction kernel appears to be very important part of the model. No individual -scalar nor vector -interactions provide quarkonium spectrum \citep{SAUBIC2012}. 

Nonrelativistic quantum mechanical limit of the scalar interaction is given by the exponential potential. Consequently the spectrum of radially excited states do not correspond with linear Regge trajectory but the gap between the states increase with the mass  of the bound states. This is in very good agreement with recent experiments.
The intercept -the gap between $J/\Psi$ and the first excited state $\Psi(2S)$
is driven by the interplay of the  strength of vectorial and scalar interaction. As we have checked numerically, coulombic term can be regarded as a perturbation only for existing energy levels for which it  slightly shifts existing energy levels. On the other side, its omission  would lead to the appearance  of new states  bellow existing 2S mass. We found that the correct adjustment  of the energy intercept is major effect of the  effective one gluon exchange interaction. The running coupling is fixed here and its numerical value differs significantly from the one known from quantum mechanical phenomenology (it is more then two times smaller). At present stage we have no simple explanation of this fact. If our results are considered seriously then it can   suggest that average of typical  square of gluon four-momenta inside quarkonium can be considerably larger then naively expected from Schroedinger equation. A possible explanation is that the soft gluon mass
is larger then the value we consider here , $\mu_g=700-800 MeV$, however QM mechanical limit has not yet been studied in such case. 
Further, comparing with the expected quantum mechanical limit
then Bethe-Salpeter string breaking scale $\mu\simeq 350MeV$  appears to be relatively larger then the one used in potential model $\mu\simeq 150MeV$ \citep{Licha2009,Chao1}.

There is another challenging aspect of this problem, as most of the SDE-BSE studies rely on the ladder truncation of the equations system.  After the inclusion of running quark masses, renormalization wave functions, the techniques can be very useful in calculation of heavy-light and light flavored meson as well.  We also expect that the knowledge of off-shell behavior (e.g. $q.P$ dependence) of the amplitudes is important in various hadronic processes.
Due to this it is worthwhile to extend our study to the more complete calculations, which open possible first principle calculation of production and decay mechanisms of heavy flawour hadrons.



\begin{thebibliography}{00}
%




\bibitem{Bugg:2004xu} 
 D.~V.~Bugg,
 Phys.\ Rept.\  {\bf 397}, 257 (2004)
arXiv:hep-ex/0412045].
%
\bibitem{Afonin:2006vi}
S.~S.~Afonin,
Phys.\ Lett.\  B {\bf 639}, 258 (2006)
[arXiv:hep-ph/0603166];
%
\bibitem{Glozman:2007ek}
  L.~Y.~Glozman,
  arXiv:hep-ph/0701081.
 %
\bibitem{fig1afon}
The large degeneracy is clear in Fig.1 of Ref \citep{Afonin:2006vi}{\em b} and in Fig. 2 of
Ref \citep{Glozman:2007ek}.
%
\bibitem{Aker:1992ny}
  E.~Aker {\it et al.}  [Crystal Barrel Collaboration],
  Nucl.\ Instrum.\ Meth.\  A {\bf 321} (1992) 69.
%
\bibitem{Wilson:1974}
K.G.Wilson, phys.Rev. D {\bf10}, 2445 (1974).
%
\bibitem{Eichten:1978}
E.Eichten, K.Gottfried, T.Kinoshite,K.D. lane and T.M. Yan, Phys. rev. D {\bf17}, 3090 (1978) [Erratum-ibid. D{\bf21},313 (1980)].
%
\bibitem{Bali2001}
G.S.Bali, Phys. Rept. {\bf343} (2001).
%
\bibitem{Greensite2003}
J. Greensite, S. Olejnik, Phys. Rev. D {\bf67}
%
\bibitem{Licha2009}
Bai-Quing Linad Kuang-Ta Chao, arXiv:0903.5506v2.
%
\bibitem{Chao1}
K.T.Chao and J.H. Liu, in Proceedings of the Workshop on Weak Interactions and CP violations, Beijing, August22-26,1989 World Scientific, Singapore (1990)
%
\bibitem{Vento2011}
P.Gonzales, V. Mathieu, V.Vento arXiv:1108.2347.
%
\bibitem{RUPP}
Eef van Beveren, George Rupp,
 Evidence for the psi(5S) and psi(4D) c-cbar vector resonances, arXiv:1005.3490.
%
\bibitem{PETR}
P. Bydzovsky, Yu.S. Surovtsev, Rho-Like Mesons from Analysis of the Pion-Pion Scattering
    , arXiv:0711.4748, talk presented at the XII Int. Conf. on Hadron Spectroscopy - Hadron07 (8-13 October 2007, Frascati, Italy).
%
\bibitem{najit}
T. Goldman and  R. Silbar, Phys. Rev. C {\bf85}, 015203 (2012). 
%
\bibitem{PBC1}
A. C. Aguilar, D. Binosi, J. Papavassiliou, arXiv:1107.3968.
%
\bibitem{PBC2}
D. Binosi, J. Papavassiliou, Phys. Rep.  479, (2009), arXiv:0909.2536 .
%
\bibitem{Mendes}
Attilio Cucchieri and  Tereza Mendes, Phys. Rev. D{\bf 81}016005, (2010).
%
\bibitem{RQM}
  A.Yu.Dubin, A.B.Kaidalov, Yu.A.Simonov,    Phys. Atom. Nucl. {\bf 56} 1745 (1993). 
  A.Yu.Dubin, A.B. Kaidalov, Yu.A. Simonov, Yad. Fiz. {\bf 56}, 213 (1993).
  A.Yu.Dubin, Phys. Lett. B {\bf323}, 41 (1994). 
  Yu.S. Kalashnikova, A.V. Nefediev, Yu.A. Simonov, Phys. Rev. D{\bf 64} 014037 (2001).
%
\bibitem{TJON1993}
Peter C. Tiemeijer and J.A. Tjon, Phys. Rev. C{\bf48} 896,(1993).
%
\bibitem{SAULI2}
V.S. arXiv:1207.2621.
%
\bibitem{INSTANT}
W. Lucha, AIP Conf. Proc. 1317, 122-127 (2011). 
for futher 3d reduction pf BSE see:
%
Kadyshevsky, Nucl. Phys. B{\bf 6} 125, (1968) .
%
Gross, Phys.Rev. {\bf186} 1448, (1969).
%
Gross, Phys. Rev. C{\bf 26} 2203, (1982). 
%
\bibitem{BETHE}
H. Bethe and E. Salpeter, {\it Phys. Rev.} {\bf 84},1232 (1951);
Y. Nambu, {\it Prog.   Theor.   Phys. } {\bf5}, 614 (1950).
%
\bibitem{JAMU}
H. J. Munczek and P. Jain, Phys. Rev. D {\bf 46}, 438 (1992);
%
P. Jain and H. J. Munczek, Phys. Rev. D {\bf 48}, 5403 (1993).
%
\bibitem{BLAKRA2010}
 M. Blank, A. Krassnigg, Comput. Phys. Commun. {\bf 182}, 1391 (2011). 
%
\bibitem{SAULI}
V. Sauli and J. Adam, Phys.Rev. D67 (2003) 085007.
%
\bibitem{ALKOFER}
R. Alkofer, S. Ahlig,	Annals Phys. {\bf 275}, 113 (1999) 
,arXiv:hep-th/9810241v2.
%
\bibitem{SAUBIC2012}
V.Sauli, P.Bicudo, in preparation.
%
\bibitem{sauli}
V.S. arXiv:1207.2621.
%
\bibitem{QUARKONIA}
N.~Brambila {\it et al.}, 
 Eur.Phys.J. C{\bf 71}, 1534 (2011), arXiv:hep-ph/1010.5827.
%










\end{thebibliography}
\end{document}